\documentclass[12pt]{article}

\usepackage[english]{babel}
\usepackage[T1]{fontenc}

\usepackage[a4paper,top=2cm,bottom=2cm,left=2cm,right=2cm,marginparwidth=1.75cm]{geometry}

\usepackage{amsmath, amsthm, amssymb}
\usepackage{graphicx}
\usepackage[natbib, maxcitenames=3, style=apa, uniquename=false, uniquelist=false]{biblatex}
\numberwithin{table}{section}

\usepackage{float}
\usepackage{setspace}

\usepackage[colorlinks=true, allcolors=blue]{hyperref}

\newcommand{\bx}{\mathbf{x}}

\newcommand{\bh}{\mathbf{h}}

\newcommand{\ba}{\mathbf{a}}
\newcommand{\bp}{\mathbf{p}}

\newcommand{\bw}{\mathbf{w}}
\newcommand{\bd}{\mathbf{d}}

\newcommand{\bv}{\mathbf{v}}
\newcommand{\bQ}{\mathbf{Q}}

\newcommand{\bbeta}{\mathbf{\beta}}

\newcommand{\btau}{\mathbf{\tau}}

\newcommand{\Exp}{\operatorname{Exp}}
\newcommand{\Ber}{\operatorname{Bernoulli}}
\newcommand{\Uni}{\operatorname{Uniform}}

\title{A note on joint calibration estimators \\for totals and quantiles}
\author{Maciej Beręsewicz\footnote{Poznań University of Economics and Business, Institute of Informatics and Quantitative Economics, Department of Statistics, Al. Niepodległości 10, 61-875 Poznań, Poland, E-mail: \url{maciej.beresewicz@ue.poznan.pl}; Statistical Office in Poznań, ul. Wojska Polskiego 27/29 60-624 Poznań, Poland}, Marcin Szymkowiak\footnote{Poznań University of Economics and Business, Institute of Informatics and Quantitative Economics, Department of Statistics, Al. Niepodległości 10, 61-875 Poznań, Poland; Statistical Office in Poznań, ul. Wojska Polskiego 27/29 60-624 Poznań, Poland}}
\date{}

\begin{document}
\maketitle

\doublespacing

\begin{abstract}
In this paper, we combine calibration for population totals proposed by \citet{deville1992calibration} with calibration for population quantiles introduced by \citet{harms2006calibration}. We also extend the pseudo-empirical likelihood method proposed by \citet*{chen2002using}. This approach extends the calibration equations for totals by adding relevant constraints on quantiles of continuous variables observed in the data. The proposed approach can be easily applied to handle non-response and data integration problems, and results in a~single vector of weights. Furthermore, it is a multipurpose solution, i.e. it makes it possible to improve estimates of means, totals and quantiles for a~variable of interest in one step. In a~limited simulation study, we compare the proposed joint approach with standard calibration, calibration using empirical likelihood and the correctly specified inverse probability weighting estimator. Open source software implementing the proposed method is available.
\end{abstract}


\clearpage

\doublespacing

\section{Introduction}

Calibration weighting is a method commonly used in survey sampling to adjust original design weights for sampled elements to reproduce known population totals for all auxiliary variables \citep{deville1992calibration}. Following the calibration paradigm, it can also be used to reproduce known population quantiles for all benchmark variables \citep{harms2006calibration}. This technique is also used in surveys to compensate for nonsampling errors, such as nonresponse or coverage errors \citep{sarndal2005estimation}. By appropriately adjusting the weights, it is not only possible to ensure consistency with known structures for key variables from other data sources, such as censuses or registers, but also to reduce the bias and improve the  precision of final estimates. 

In this article, we propose a~joint calibration approach to estimate the total or quantile of order $\alpha$ for the variable of interest $y$. Final calibration weights $w_{k}$ reproduce known population totals and quantiles for all auxiliary variables. At the same time, they help to reduce the bias and improve the precision of estimates. The proposed method is based on the classic approach to calibration and simultaneously takes into account calibration equations for totals and quantiles of all auxiliary variables.

The paper has the following structure. In section \ref{sec-basic} we introduce the basic setup: notation, calibration for totals and quantiles. In section \ref{sec-proposed} we describe the procedure of joint calibration of totals and quantiles. In section \ref{sec-sims} we describe results of a simulation study based on \citet{chen2020doubly}. Section \ref{sec-summary} summarises the paper.

\section{Theoretical basis}\label{sec-basic}
\subsection{Calibration estimator for a total}

In most applications the goal is to estimate a~finite population total $\tau_{y}=\sum_{k\in U}y_{k}$ or the mean $\bar{\tau}_{y}=\tau_{y}/N$ of the variable of interest $y$, where $U$ is the population of size $N$. The well-known estimator of a finite population total is the Horvitz-Thompson estimator, which is expressed as $\hat{\tau}_{y\pi}=\sum_{k=1}^{n}d_{k}y_{k}=\sum_{k\in s}{d_{k}y_{k}}$, where $s$ denotes a~probability sample of size $n$, $d_{k}=1/\pi_{k}$ is a~design weight and $\pi_{k}$ is the first-order inclusion probability of the $i$-th element of the population $U$. This estimator is unbiased for $\tau_{Y}$ i.e. $E\left(\hat{\tau}_{y\pi}\right)=\tau_{Y}$. 

Let $\bx_{k}^{\circ}$ be a~$J_{1}$-dimensional vector of auxiliary variables (benchmark variables) for which $\tau_{\bx}=\sum_{k\in U}\bx_{k}^{\circ}=\left(\sum_{k\in U}x_{k1},\ldots,\sum_{k\in U}x_{kJ_{1}}\right)^T$ is assumed to be known. In most cases in practice the $d_{k}$ weights do not reproduce known population totals for benchmark variables $\bx_{k}^{\circ}$. It means that the resulting estimate $\hat{\tau}_{\bx\pi}=\sum_{k\in s}{d_{k}\bx_{k}^{\circ}}$ is not equal to $\tau_{\bx}$. The main idea of calibration is to look for new calibration weights $w_{k}$ which are as close as possible to original design weights $d_{k}$ and reproduce known population totals $\tau_{\bx}$ exactly. In other words, in order to find new calibration weights $w_{k}$ we have to minimise a~distance function $D\left(\bd,\bv\right)=\sum _{k\in s}d_{k}\hspace{2pt} G\hspace{0pt}\left(\frac{v_{k}}{d_{k}}\right) \to \textrm{min}$ to fulfil calibration equations $\sum_{k\in s}v_{k}\bx_{k}^{\circ} = \sum_{k\in U}\bx_{k}^{\circ}$, where $\bd=\left(d_{1},\ldots,d_{n}\right)^T$, $\bv=\left(v_{1},\ldots,v_{n}\right)^T$ and $G\left(\cdot\right)$ is a~function which must satisfy some regularity conditions: $G\left(\cdot\right)$ is strictly convex and twice continuously differentiable, $G\left(\cdot\right)\geq 0$, $G\left(1\right)=0$, $G'\left(1\right)=0$ and $G''\left(1\right)=1$. Examples of $G\left(\cdot\right)$ functions are given by \citet{deville1992calibration}. For instance, if $G\left(x\right)=\frac{\left(x-1\right)^{2}}{2}$, then using the method of Lagrange multipliers the final calibration weights $w_{k}$ can be expressed as $w_{k}=d_{k}+d_{k}\left(\tau_{\bx}-\hat{\tau}_{\bx\pi}\right)^T\left(\sum_{j\in s}d_{j}\bx_{j}^{\circ}\bx_{j}^{\circ T}\right)^{-1}\bx_{k}^{\circ}$. It is worth adding that in order to avoid negative or large $w_{k}$ weights in the process of minimising the $D\left(\cdot\right)$ function, one can consider some boundary constraints $L\leq \frac{w_{k}}{d_{k}}\leq U$, where $\ 0\leq L\leq 1 \leq U,\  k=1,\ldots,n$. The final calibration estimator of a population total $\tau_{y}$ can be expressed as $\hat{\tau}_{y\bx}=\sum_{k\in s}w_{k}y_{k}$, where $w_{k}$ are calibration weights obtained under a~specific chosen $G\left(\cdot\right)$ function.

\subsection{Calibration estimator for a quantile}

\citet{harms2006calibration} considered the estimation of quantiles using the calibration approach in a~way very similar to the what \citet{deville1992calibration} proposed for a~finite population total $\tau_{y}$. By analogy, in their approach it is not necessary to know values for all auxiliary variables for all units in the population. It is enough to know the corresponding quantiles for the benchmark variables. We will briefly discuss the~problem of finding calibration weights in this setup.  

We want to estimate a~quantile $Q_{y,\alpha}$ of order $\alpha \in \left(0,1\right)$ of the variable of interest $y$, which can be expressed as $Q_{y,\alpha}=\mathrm{inf}\left\{t\left|F_{y}\left(t\right)\geq \alpha \right.\right\}$, where $F_{y}\left(t\right)=N^{-1}\sum_{k\in U}H\left(t-y_{k}\right)$ and the Heavyside function is given by 
\begin{equation}\label{H}
H\left(t-y_{k}\right)=\left\{ \begin{array}{ll}
1, & \ t \geq y_{k},\\
0, & \ t<y_{k}.\\
\end{array} \right.
\end{equation}

We assume that $\bQ_{\bx,\alpha}=\left(Q_{x_{1},\alpha},\ldots,Q_{x_{J_{2}},\alpha}\right)^{T}$ is a~vector of known population quantiles of order $\alpha$ for a vector of auxiliary variables $\bx_{k}^{*}$, where $\alpha \in \left(0,1\right)$ and $\bx_{k}^{*}$ is a~$J_{2}$-dimensional vector of auxiliary variables. It is worth noting that, in general, the numbers $J_{1}$ and $J_{2}$ of the auxiliary variables are different. It may happen that for a~specific auxiliary variable its population total and the corresponding quantile of order $\alpha$ will be known. However, in most cases quantiles will be known for continuous auxiliary variables, unlike totals, which will generally be known for categorical variables. In order to find new calibration weights $w_{k}$ which reproduce known population quantiles in a~vector $Q_{\bx,\alpha}$, an interpolated distribution function estimator of $F_{y}\left(t\right)$ is defined as $
\hat{F}_{y,cal}(t)=\frac{\sum_{k \in s} w_{k} H_{y, s}\left(t, y_{k}\right)}{\sum_{k \in s} w_{k}} 
$, where the Heavyside function in formula (\ref{H}) is replaced by the modified function $H_{y, s}\left(t, y_{k}\right)$ given by

\begin{equation}
H_{y, s}\left(t, y_{k}\right)=\left\{
\begin{array}{ll}
1, & y_{k} \leqslant L_{y, s}(t), \\ 
\beta_{y, s}\left(t\right), & y_{k}=U_{y, s}\left(t\right), \\ 
0, & y_{k}>U_{y, s}\left(t\right),
\end{array}\right.
\end{equation}

\noindent where $L_{y, s}\left(t\right)=\max \left\{\left\{y_{k}, k \in s \mid y_{k} \leqslant t\right\} \cup\{-\infty\}\right\}$, $U_{y, s}\left(t\right)=\min \left\{\left\{y_{k}, k \in s \mid y_{k}>t\right\} \cup\{\infty\}\right\}$ and $\beta_{y, s}\left(t\right)=\frac{t-L_{y, s}\left(t\right)}{U_{y, s}\left(t\right)-L_{y, s}\left(t\right)}$ for $k=1,\ldots,n$, $t \in \mathbb{R}$. A calibration estimator of quantile $Q_{y,\alpha}$ of order $\alpha$ for variable $y$ is defined as $\hat{Q}_{y,cal,\alpha}=\hat{F}_{y,cal}^{-1}(\alpha)$, where a~vector $\bw=\left(w_{1},\ldots,w_{n}\right)^{T}$ is a~solution of optimization problem $D\left(\bd,\bv\right)=\sum _{k\in s}d_{k}\hspace{2pt} G\hspace{0pt}\left(\frac{v_{k}}{d_{k}}\right) \to \textrm{min}$ subject to the calibration constraints $\sum_{k\in s}v_{k}=N$ and  $\hat{\bQ}_{\bx,cal,\alpha}=\left(\hat{Q}_{x_{1},cal,\alpha},\ldots,\hat{Q}_{x_{J_{2}},cal,\alpha}\right)^{T}=\bQ_{\bx,\alpha}$ or equivalently $\hat{F}_{x_{j},cal}\left(Q_{x_{j},\alpha}\right)=\alpha$, where $j=1,\ldots,J_{2}$.

As in the previous case, if $G\left(x\right)=\frac{\left(x-1\right)^{2}}{2}$ then using the method of Lagrange multipliers the final calibration weights $w_{k}$ can be expressed as $w_{k}=d_{k}+d_{k}\left(\mathbf{T_{a}}-\sum_{k\in s}{d_{k}\ba_{k}}\right)^{T}\left(\sum_{j\in s}{d_{j}}\ba_{j}\ba_{j}^{T}\right)^{-1}\ba_{k}$, where $\mathbf{T_{a}}=\left(N,\alpha,\ldots,\alpha\right)^{T}$ and the elements of $\ba_{k}=\left(1,a_{k1},\ldots,a_{kJ_{2}}\right)^{T}$ are given by
\begin{equation}
a_{kj}=\left\{\begin{array}{lll} 
N^{-1},& \quad x_{kj}\leq L_{x_{j},s}\left(Q_{x_{j},\alpha}\right),\\
N^{-1}\beta_{x_{j},s}\left(Q_{x_{j},\alpha}\right), & \quad x_{kj}=U_{x_{j},s}\left(Q_{x_{j},\alpha}\right),\\
0,& \quad x_{kj}> U_{x_{j},s}\left(Q_{x_{j},\alpha}\right),\\
\end{array} \right.
\end{equation}
with $j=1,\ldots,J_{2}$.

Assuming that $y_{1}\leq y_{2}\ldots \leq y_{n}$ it can be shown that if there exists $p\!\in\!\!{\left\{1,\ldots,n-1\right\}}$ such that $\hat{F}_{y,cal}\left(y_{p}\right)\leq \alpha$, $\hat{F}_{y,cal}\left(y_{p+1}\right)> \alpha$ and $\hat{F}_{y,cal}$ is invertible at point $\hat{Q}_{y,cal,\alpha}$
then the calibration estimator $\hat{Q}_{y,cal,\alpha}$ of quantile $Q_{y,\alpha}$ of order $\alpha \in \left(0,1\right)$ can be expressed as $\hat{Q}_{y,cal,\alpha}=y_{p}+\frac{N\alpha-\sum_{i=1}^{p}{w_{i}}}{w_{p+1}}\left(y_{p+1}-y_{p}\right)$.

\section{Proposed method}\label{sec-proposed}

We propose a~simple method that jointly calibrates weights for totals and quantiles. The resulting calibrated weights $w_{k}$ will allow us to retrieve known population totals and quantiles of auxiliary variables simultaneously. In the case of a~single scalar auxiliary variable $x$, the final calibration estimator based on weights $w_{k}$ delivers an exact population total and quantile for variable $y$ when the relationship between $y$ and $x$ is exactly linear i.e. when $y_{k}=\beta x_{k}$ for all $k\in U$.

Let us assume that we are interested in estimating a population total $\tau_{y}$ and/or quantile $Q_{y,\alpha}$ of order $\alpha$, where $\alpha \in \left(0,1\right)$ for variable of interest $y$. Let $\bx_{k}=\left(\begin{smallmatrix}\bx_{k}^{\circ}\\1\\\bx_{k}^{*}\end{smallmatrix}\right)$ be a~$J+1$-dimensional vector of auxiliary variables, where $J=J_{1}+J_{2}$. We assume that for $J_{1}$ variables a~vector of population totals $\tau_{\bx}$ is known and for $J_{2}$ variables a~vector $\bQ_{\bx,\alpha}$ of population quantiles is known. In practice it may happen that for the same auxiliary variable we know its population total and quantile. We do not require that the complete auxiliary information described by the vector $\bx_{k}$ is known for all $k\in U$; however, for some auxiliary variables unit-population data would be necessary, because accurate quantiles are not likely to be known from other sources \citep{sarndal2007calibration}. 

Our main aim is to find new calibration weights $w_{k}$ which are as close as possible to original design weights $d_{k}$ and for some auxiliary variables reproduce known population totals and for the remaining benchmark variables -- reproduce known population quantiles exactly. In our joint approach we are looking for a~vector $\bw=\left(w_{1},\ldots,w_{n}\right)^{T}$ which is a~solution of the optimization problem $D\left(\bd,\bv\right)=\sum _{k\in s}d_{k}\hspace{2pt} G\hspace{0pt}\left(\frac{v_{k}}{d_{k}}\right) \to \textrm{min}$ subject to the calibration constraints $\sum_{k\in s}v_{k}=N$, $\sum_{k\in s}v_{k}\bx_{k}^{\circ} = \tau_{\bx}$ and  $\hat{\bQ}_{\bx,cal,\alpha}=\bQ_{\bx,\alpha}$. In general, $J+1$ calibration equations have to be fulfilled. Alternatively, calibration equations can be expressed as $\sum_{k\in s}v_{k}\bx_{k}^{\circ} = \tau_{\bx}$ and $\sum_{k\in s}v_{k}\ba_{k}=\mathbf{T_{a}}$ with possible boundary constraints on calibration weights. 

Assuming a~quadratic metric $D\left(\cdot\right)$, which is based on $G\left(x\right)=\frac{\left(x-1\right)^{2}}{2}$ function, an explicit solution of the above optimization problem can be derived. This solution is similar to the calibration weights for totals and quantiles. Let $\bh_{\bx}=\binom{\tau_{\bx}}{\mathbf{T_{a}}}$ and $\hat{\bh}_{\bx}=\binom{\sum_{k\in s}d_{k}\bx_{k}^{\circ}}{\sum_{k\in s}d_{k}\ba_{k}}$. Then the vector of calibration weights $\bw=\left(w_{1},\ldots,w_{n}\right)^{T}$ which solves the above optimization problem satisfies the relation:
\begin{equation}
w_{k}=d_{k}+d_{k}\left(\bh_{\bx}-\hat{\bh}_{\bx}\right)^{T}\left(\sum_{j\in s}{d_{j}}\bx_{j}\bx_{j}^{T}\right)^{-1}\bx_{k}.
\end{equation}

\noindent \textbf{Remark 1}: In our proposed method we assume that we reproduce known population totals and known population quantiles for a~set of benchmark variables simultaneously. This approach can be easily extended by assuming that for estimated totals or quantiles are reproduced for some auxiliary variables. Moreover, we assumed that the process of calibration is based on a~particular quantile (of order $\alpha$). For instance, it could be a median i.e. $\alpha=0.5$. Nothing stands in the way of searching for calibration weights which reproduce population totals and a~set of population quantiles (for example quartiles) for a~chosen set of auxiliary variables. Moreover, the proposed method can be easily extended for the generalized calibration, in particular, for not missing at random non-response \citep{kott2010using}, or empirical likelihood by adding additional constraints on quantiles, i.e. $\sum_{k \in s} p_{k}a_{kj} = \frac{\alpha}{N}$, where $j=1,\ldots,J_{2}$ and $p_{k}$ are elements of the vector $\bp=\left(p_{1},\ldots,p_{n}\right)^{T}$ which is is a~discrete probability measure over the sample $s$ \citep{wu2020sampling}. 

\noindent \textbf{Remark 2}: In our approach we use an interpolated distribution function $H_{y,s}$, which is a~simple modification of the Heavyside function defined in (\ref{H}). From a~practical point of view a~smooth approximation to the step function, based on the logistic function can be used i.e. $H\left(x\right)\approx\frac{1}{2}+\frac{1}{2}\tanh{kx}=\frac{1}{1+e^{-2kx}}$, where a~larger value of $k$ corresponds to a~sharper transition at $x = 0$. 

\noindent \textbf{Remark 3}: The proposed method can be easily applied to data from household surveys, where integrated calibration is applied, i.e. when the weights of particular household members should be equal. This can be in particularly useful in the case of EU-SILC, where information from administrative data can be used to provide population distributions for auxiliary variables.



\section{Simulation study based on \citet{chen2020doubly}}\label{sec-sims}

In this section, we present results of a~design-based simulation study using \citet{chen2020doubly} setting. Our population $U$ is equal to $N=20,000$, and the observed set (non-probability sample), which suffers from selection bias, is equal to $n=10,000$ . We generate auxiliary variables $x_{k1}=z_{k1}$, $x_{k2}=z_{k2}+0.3 x_{k1}$, $x_{k3}=z_{k3}+0.2(x_{k1}+x_{k2})$ and $x_{k4}=z_{k4}+0.1(x_{k1}+x_{k2}+x_{k3})$ where $z_{k1} \sim \Ber(0.5)$, $z_{k2} \sim \Uni(0,2)$, $z_{k3} \sim \Exp(1)$ and $z_{k4} \sim \chi^{2}(4)$ for all $k\in U$. The target variable $y$ is generated using the following linear model

$$
y_{k}=2+x_{k1}+x_{k2}+x_{k3}+x_{k4}+\sigma \varepsilon_{k}, \quad k=1,2, \ldots, N,
$$

\noindent where $\varepsilon_{i}$'s are independent and identically distributed random variables (i.i.d.) from normal distribution $\mathcal{N}(0,1)$. Values of $\sigma$ are chosen such that the correlation coefficient $\rho$ between $y$ and the linear predictor $\bx^T \bbeta$ is controlled at $0.3, 0.5$ and $0.8$ (such variables $y$ are denoted as $y_1, y_2$, and $y_3$ respectively). The true propensity scores $\pi_{k}$ for the non-probability sample follow the logistic regression model:

$$
\log \left(\frac{\pi_{k}}{1-\pi_{k}}\right)=\theta_{0}+0.1 x_{k1}+0.2 x_{k2}+0.1 x_{k3}+0.2 x_{k4},
$$

\noindent where $\theta_{0}$ is chosen such that $\sum_{k\in U} \pi_{k}=n$. The non-probability sample is selected by the Poisson sampling method with inclusion probabilities specified by $\pi_{k}$ and the target sample size $n$. 

We conducted a~simulation study where population totals were known for all auxiliary variables $x_{1}-x_{4}$, while population quartiles $Q_{1}$, $Q_{2}$ and deciles $D_{1},\ldots,D_{9}$ were known for $x_{2}-x_{4}$ only. For all calibration estimators, we used raking (exponential function) to ensure positive weights:

\begin{itemize}
    \item $\hat{\mu}_{\text{Naive}}$ -- sample mean or quantile based on a~sample $n$,
    \item $\hat{\mu}_{IPW}$ -- we used all auxiliary variables assuming logistic regression with calibration constraints (as in \citet[sec 3.1]{chen2020doubly}) ,
    \item $\hat{\mu}_{CAL}$ -- calibration using population size $N$ and totals for all benchmark variables i.e. $\btau_{\bx} = (N, \tau_{x_1}, \tau_{x_2}, \tau_{x_3}, \tau_{x_4})^T$,
    \item $\hat{\mu}_{QCAL1}$ -- calibration using population quartiles of order 0.25 and 0.75 and population deciles i.e. quantiles of order 0.1 to 0.9 for auxiliary variables $x_{2}, x_{3}, x_{4}$, and population size $N$,
    \item $\hat{\mu}_{QCAL2}$ -- calibration using population size and totals for variables $x_{1}-x_{4}$, population quartiles of order 0.25 and 0.75 and deciles for auxiliary variables $x_{2}-x_{4}$,
    \item $\hat{\mu}_{EL}$ -- calibration using empirical likelihood (EL) using constraints on means ($\sum_{k \in s} p_k\bx_k^{\circ} = \frac{\btau_{\bx}}{N}$) of auxiliary variables $x_{1}-x_{4}$,
    \item $\hat{\mu}_{QEL1}$ -- calibration using EL using constraints on quartiles of order 0.25 and 0.75 and deciles ($\sum_{k \in s} p_ka_{kj} = \frac{\alpha}{N}$) of auxiliary variables $x_{2}-x_{4}$,
    \item $\hat{\mu}_{QEL2}$ -- calibration using  EL combining constraints described by $\hat{\mu}_{EL}$ and  $\hat{\mu}_{QEL1}$.
\end{itemize}

In the results, we report bias, standard error (SE) and relative mean square error (RMSE)  based on $R=1,000$ Monte Carlo simulations for each $y_1, y_2$ and $y_3$ variables: $\text{Bias}  = \bar{\hat{\mu}} - \mu$, $ \text{SE}= \sqrt{\frac{\sum_{r=1}^R \left(\hat{\mu}^{(r)} - \bar{\hat{\mu}}\right)^2}{R-1}}$ and $\text{RMSE} = \sqrt{ \text{Bias}^2 + \text{SE}^2}$, where $\bar{\hat{\mu}} = \sum_{r=1}^{R}\hat{\mu}^{(r)} / R$ and $\hat{\mu^{\left(r\right)}}$ is an estimate of mean/quantile in $r$-th replication. We present results of our simulation study in Table \ref{tab-results-sim} for the calibration estimator of the mean and quartiles of variable of interest $y$ (for three variants).

\begin{table}[ht!]
\centering
\small
\caption{Simulation results: bias, SE and RMSE for $y_1$, $y_2$ and $y_3$ (all numbers multiplied by 100)}
\label{tab-results-sim}
\begin{tabular}{lrrrrrrrrr}
  \hline
Estimator & \multicolumn{3}{c}{$y_1 (\rho=0.3)$} & \multicolumn{3}{c}{$y_2 (\rho=0.5)$}  & \multicolumn{3}{c}{$y_3 (\rho=0.8)$}  \\ 
 &  Bias & SE & RMSE & Bias & SE & RMSE & Bias & SE & RMSE \\ 
  \hline 
  \multicolumn{10}{c}{Mean} \\
  \hline
  $\hat{\mu}_{\text{Naive}}$ & 94.58 & 7.13 & 94.85 & 93.92 & 4.21 & 94.02 & 93.48 & 2.54 & 93.52 \\ 
  $\hat{\mu}_{IPW}$ & -0.17 & 7.86 & 7.86 & -0.09 & 4.26 & 4.26 & -0.04 & 1.84 & 1.84 \\ 
  $\hat{\mu}_{CAL}$ & 1.10 & 7.74 & 7.82 & 0.60 & 4.20 & 4.24 & 0.26 & 1.81 & 1.83 \\ 
  $\hat{\mu}_{QCAL1}$ & 3.75 & 7.84 & 8.69 & 4.04 & 4.28 & 5.88 & 4.24 & 1.90 & 4.64 \\ 
  $\hat{\mu}_{QCAL2}$ & 0.12 & 7.85 & 7.85 & 0.07 & 4.25 & 4.25 & 0.03 & 1.84 & 1.84 \\ 
  $\hat{\mu}_{EL}$ & 0.06 & 7.82 & 7.82 & 0.03 & 4.24 & 4.24 & 0.01 & 1.83 & 1.83 \\ 
  $\hat{\mu}_{QEL1}$ & 3.37 & 7.88 & 8.57 & 3.84 & 4.29 & 5.76 & 4.16 & 1.91 & 4.58 \\
  $\hat{\mu}_{QEL2}$ & -0.24 & 7.88 & 7.89 & -0.13 & 4.28 & 4.28 & -0.06 & 1.85 & 1.85 \\ 
  \hline 
  \multicolumn{10}{c}{1$^{th}$ Quartile (25\%)} \\
  \hline
  $\hat{\mu}_{\text{Naive}}$ & 94.52 & 11.24 & 95.18 & 89.55 & 6.82 & 89.81 & 79.04 & 3.37 & 79.11 \\ 
  $\hat{\mu}_{IPW}$ & 0.35 & 12.16 & 12.16 & -0.14 & 6.89 & 6.89 & -0.17 & 3.62 & 3.62 \\ 
  $\hat{\mu}_{CAL}$ & 5.06 & 12.20 & 13.21 & 6.27 & 6.56 & 9.08 & 9.35 & 3.89 & 10.13 \\ 
  $\hat{\mu}_{QCAL1}$ & 4.91 & 12.37 & 13.31 & 4.21 & 6.73 & 7.94 & 4.84 & 3.77 & 6.14 \\ 
  $\hat{\mu}_{QCAL2}$ & 2.20 & 12.21 & 12.40 & 2.31 & 6.80 & 7.18 & 3.31 & 3.64 & 4.92 \\ 
  $\hat{\mu}_{EL}$ & 1.34 & 12.17 & 12.24 & 1.35 & 6.84 & 6.97 & 1.86 & 3.61 & 4.06 \\ 
  $\hat{\mu}_{QEL1}$ & 3.45 & 12.38 & 12.85 & 2.51 & 6.84 & 7.29 & 2.29 & 3.68 & 4.34 \\ 
  $\hat{\mu}_{QEL2}$ & 0.68 & 12.24 & 12.26 & 0.30 & 6.90 & 6.90 & 0.44 & 3.59 & 3.62 \\
 \hline 
  \multicolumn{10}{c}{2$^{th}$ Quartile (50\%)} \\
  \hline
  $\hat{\mu}_{\text{Naive}}$ & 90.01 & 8.95 & 90.45 & 95.27 & 5.25 & 95.41 & 92.23 & 2.82 & 92.27 \\ 
  $\hat{\mu}_{IPW}$ & -0.46 & 9.36 & 9.37 & -0.34 & 5.07 & 5.08 & -0.14 & 3.17 & 3.17 \\ 
  $\hat{\mu}_{CAL}$ & 2.54 & 9.19 & 9.53 & 2.75 & 4.94 & 5.65 & 5.67 & 2.82 & 6.34 \\ 
  $\hat{\mu}_{QCAL1}$  & 3.55 & 9.43 & 10.07 & 3.49 & 5.06 & 6.15 & 4.18 & 2.91 & 5.09 \\ 
  $\hat{\mu}_{QCAL2}$ & 0.92 & 9.38 & 9.42 & 0.95 & 5.00 & 5.09 & 1.99 & 3.01 & 3.61 \\ 
  $\hat{\mu}_{EL}$ & 0.39 & 9.27 & 9.28 & 0.80 & 4.99 & 5.06 & 2.31 & 2.99 & 3.78 \\ 
  $\hat{\mu}_{QEL1}$ & 2.52 & 9.40 & 9.73 & 2.93 & 5.06 & 5.85 & 3.43 & 2.96 & 4.53 \\ 
  $\hat{\mu}_{QEL2}$ & -0.17 & 9.40 & 9.40 & 0.19 & 5.05 & 5.05 & 0.99 & 3.08 & 3.23 \\
  \hline 
  \multicolumn{10}{c}{3$^{th}$ Quartile (75\%)} \\
  \hline
  $\hat{\mu}_{\text{Naive}}$ & 94.56 & 9.34 & 95.02 & 104.29 & 5.60 & 104.44 & 104.51 & 3.96 & 104.58 \\ 
  $\hat{\mu}_{IPW}$ & -1.24 & 11.18 & 11.24 & -0.44 & 5.78 & 5.80 & 0.09 & 3.13 & 3.14 \\ 
  $\hat{\mu}_{CAL}$ & -1.86 & 11.27 & 11.42 & -2.58 & 5.74 & 6.29 & -2.20 & 3.02 & 3.74 \\ 
  $\hat{\mu}_{QCAL1}$ & 2.62 & 10.83 & 11.14 & 3.18 & 5.74 & 6.56 & 2.86 & 3.10 & 4.22 \\ 
  $\hat{\mu}_{QCAL2}$  & -1.47 & 11.24 & 11.34 & -1.22 & 5.78 & 5.91 & -1.05 & 3.00 & 3.18 \\ 
  $\hat{\mu}_{EL}$ & -1.19 & 11.19 & 11.25 & -0.46 & 5.75 & 5.77 & 0.97 & 3.16 & 3.31 \\ 
  $\hat{\mu}_{QEL1}$ & 2.84 & 10.78 & 11.15 & 4.05 & 5.70 & 6.99 & 4.64 & 2.89 & 5.47 \\ 
  $\hat{\mu}_{QEL2}$ & -1.28 & 11.21 & 11.28 & -0.30 & 5.76 & 5.77 & 0.79 & 3.07 & 3.17 \\ 
   \hline
\end{tabular}
\end{table}

The simulation results can be summarized as follows. 
\begin{enumerate}
    \item All considered calibration estimators, as expected, yield better results in terms of bias compared to $\hat{\mu}_{\text{Naive}}$. As regards the estimated mean, one can see that $\hat{\mu}_{QCAL2}$, $\hat{\mu}_{EL}$ and $\hat{\mu}_{QEL2}$ are nearly unbiased for all cases.
    \item Of all calibration estimators based on the classical  calibration paradigm ($\hat{\mu}_{QCAL}$, $\hat{\mu}_{QCAL1}$ and $\hat{\mu}_{QCAL2}$) the proposed joint calibration estimator $\hat{\mu}_{QCAL2}$ has the best statistical properties. The simultaneous consideration of population totals and quantiles for auxiliary variables contributes to the reduction of the estimator bias. 
    \item All considered estimators yield similar results in terms of the efficiency measured by SE. The standard errors decrease as the correlation coefficient $\rho$ increases.
    \item The proposed joint approach is nearly as efficient as the correctly specified IPW estimator. 
\end{enumerate}

\section{Summary}\label{sec-summary}

In this article we have presented a simple and joint calibration estimator for totals and quantiles based on the calibration paradigm. More  precisely, using a set of auxiliary variables we are able to create a single vector of calibration weights which reproduce known population totals for some benchmark variables and known population quantiles for other variables exactly. This vector can be used to directly find a total and a quantile of the variable of interest $y$. 

In a small simulation study we compared the calibration estimator of the mean and quartiles of variable $y$ based on calibration weights obtained under different settings with other estimators using pseudo-empirical likelihood and inverse probability weighting methods. The proposed estimator based on a single set of calibration weights which reproduce known population totals and reproduce exactly quantiles for auxiliary variables performed well in our simulation. By combining information about population totals and population quantiles for auxiliary variables in a process of calibration, it is possible to decrease the~bias and RMSE in comparison to the classical estimator of a total and a quantile, where calibration equations are fulfilled for population totals and population quantiles separately. 

\section*{Acknowledgements}
The authors' work has been financed by the National Science Centre in Poland, OPUS 22, grant no. 2020/39/B/HS4/00941.

Codes to reproduce the results are freely available from the github repository: \url{https://github.com/ncn-foreigners/paper-note-joint-calibration}. An R package that implements joint calibration is available at \url{https://github.com/ncn-foreigners/jointCalib}. The package is based on calibration implemented in \texttt{survey}, \texttt{sampling} or \texttt{laeken} packages.

\printbibliography

\end{document}